\documentclass[useAMS,usenatbib]{mn2e}

\usepackage{graphicx}
\title[Eight new late-T dwarfs in the UKIDSS LAS DR1]{
Eight new T4.5--T7.5 dwarfs discovered in the UKIDSS Large Area Survey
Data Release 1
\thanks{Based on observations made with the United Kingdom Infrared
Telescope, operated by the Joint Astronomy Centre on behalf of the
U.K. Particle Physics and Astronomy Research Council.}}
\author[N. Lodieu et al.]{N. Lodieu$^{1,2}$\thanks{E-mail: nl41@star.le.ac.uk},
D. J. Pinfield$^{3}$, S. K. Leggett$^{4}$, R. F. Jameson$^{2}$, 
D. J. Mortlock$^{5}$,
\newauthor
S. J. Warren$^{5}$, B. Burningham$^{3}$, P. W. Lucas$^{3}$, K. Chiu$^{6}$, 
M. C. Liu$^{7}$, B. P. Venemans$^{8}$,
\newauthor
R. G. McMahon$^{8}$, F. Allard$^{9}$, 
I. Baraffe$^{9}$, D. Barrado y Navascu\'es$^{10}$, G. Carraro$^{11}$, 
\newauthor
S. L. Casewell$^{2}$, G. Chabrier$^{8}$, R. J. Chappelle$^{12}$, 
F. Clarke$^{13}$, A. Day-Jones$^{3}$,
\newauthor
N. R. Deacon$^{14}$, P. D. Dobbie$^{15}$, S. L. Folkes$^{3}$, 
N. C. Hambly$^{16}$, P. C. Hewett$^{8}$,
\newauthor
S. T. Hodgkin$^{8}$, H. R. A. Jones$^{3}$, T. R. Kendall$^{3}$, 
A. Magazz\`u$^{17}$, E. L. Martin$^{1}$,
\newauthor
M. J. McCaughrean$^{6}$, T. Nakajima$^{18}$, Y. Pavlenko$^{19}$, 
M. Tamura$^{18}$, C. G. Tinney$^{15}$,
\newauthor
M. R. Zapatero Osorio$^{1}$ \\
$^{1}$Instituto de Astrof\'isica de Canarias, V\'ia L\'actea s/n, 
E-38205 La Laguna, Tenerife, Spain \\
$^{2}$Department of Physics and Astronomy, University of Leicester,
University Road, Leicester LE1 7RH, U.K. \\
$^{3}$Centre for Astrophysics Research, Science and Technology Research 
Institute, University of Hertfordshire, Hatfield AL10 9AB \\
$^{4}$Gemini Observatory, 670 N. A'ohoku Place, Hilo, HI 96720, USA \\
$^{5}$Astrophysics Group, Imperial College London, Blackett Laboratory,
Prince Consort Road, London, SW7 2AZ, U.K. \\
$^{6}$School of Physics, University of Exeter, Stocker Road, 
Exeter EX4 4QL, Devon, U.K. \\
$^{7}$Institute for Astronomy, University of Hawaii, 2680 Woodlawn Drive,
Honolulu, HI 96822; Alfred P. Sloan Research Fellow \\
$^{8}$Institute of Astronomy, Madingley Road, Cambridge CB3 0HA, UK \\
$^{9}$CRAL, Ecole Normale Sup\'erieure de Lyon, 46 All\'ee d'Italie, 
F--69364,  Universit\'e de Lyon \\
$^{10}$Laboratorio de Astrof\'isica Espacial y F\'isica Fundamental,
INTA, P.O. Box 50727, E--2808 Madrid, Spain \\
$^{11}$Departamento de Astronomia, Universidad de Chile, Casilla 36-D,
Santiago, Chile \\
$^{12}$Astronomical Institute, Academy of Sciences of the Czech Republic, 
Bocni II/1401a, 141 31 Prague, Czech Republic \\
$^{13}$Department of Physics, University of Oxford, Clarendon Laboratory,
Parks Road, Oxford OX1 3PU, U.K. \\
$^{14}$Department of Astrophysics, Radboud University Nijmegen,
P.O. Box 9010, 6500 GL Nijmegen, The Netherlands \\
$^{15}$Anglo-Australian Observatory, P.O. Box 296, Epping 1710, 
Australia \\
$^{16}$Scottish Universities' Physics Alliance (SUPA),
Institute for Astronomy, School of Physics, University of Edinburgh, \\
Royal Observatory, Blackford Hill, Edinburgh EH9 3HJ, U.K.  \\
$^{17}$Fundaci\'on Galileo Galilei-INAF, Apartado 565, E-38700 
Santa Cruz de La Palma, Spain \\
$^{18}$National Astronomical Observatory, Mitaka, Tokyo 181-8588, Japan \\
$^{19}$Main Astronomical Observatory, National Academy of Sciences, 
Zabolotnoho 27, Kyiv-127 03680, Ukraine
}

\begin{document}

\date{Accepted \today. Received \today; in original form \today}

\pagerange{\pageref{firstpage}--\pageref{lastpage}} \pubyear{2005}

\maketitle

\label{firstpage}

%
%%%%%%%%%%%%%%%%%%%%%%%%%%%%%%%%%%%%%%%%%%
%%%%%%%%  Abstract  %%%%%%%%
%%%%%%%%%%%%%%%%%%%%%%%%%%%%%%%%%%%%%%%%%%
%
\begin{abstract}
We present eight new T4.5--T7.5 dwarfs identified in 
the UKIRT Infrared Deep Sky Survey (UKIDSS) Large Area Survey (LAS)
Data Release 1 (DR1). In addition we have recovered the T4.5 dwarf
SDSS J020742.91$+$000056.2 and the T8.5 dwarf
ULAS J003402.77$-$005206.7.
Photometric candidates were picked up in two-colour diagrams 
over 190 deg$^2$ (DR1) and selected in at least two filters.
All candidates exhibit near-infrared spectra with strong methane 
and water absorption bands characteristic of T dwarfs and the
derived spectral types follow the unified scheme of 
Burgasser et al.\ (2006).
We have found 6 new  T4.5--T5.5 dwarfs, one T7 dwarf, one T7.5 dwarf,
and recovered a T4.5 dwarf and a T8.5 dwarf. 
We provide distance estimates which
lie in the 15--85 pc range; the T7.5 and T8.5 dwarfs
are probably within 25 pc of the Sun. 
We conclude with a discussion of the number of T dwarfs expected after 
completion of the LAS, comparing these initial results 
to theoretical simulations. 
\end{abstract}


%
%%%%%%%%%%%%%%%%%%%%%%%%%%%%%%%%%%%%%%%%%%
%%%%%%%%  Introduction  %%%%%%%%
%%%%%%%%%%%%%%%%%%%%%%%%%%%%%%%%%%%%%%%%%%
%
\section{Introduction}
\label{dT:intro}

The advent of large-scale sky surveys has revolutionised our knowledge
of ultracool dwarfs (defined here as dwarfs with spectral types
later than M7). The first spectroscopic brown dwarfs were confirmed
in 1995: Gl229B, a T dwarf orbiting an M dwarf \citep{nakajima95}
and Teide 1 in the Pleiades open cluster \citep{rebolo95}.
Ten years on, about 500 L dwarfs with effective temperatures 
(T$_{\rm eff}$) between $\sim$2200 and $\sim$1400\,K 
\citep{basri00,leggett00} have now been 
identified, along with around 100 T dwarfs with lower temperatures 
\citep[T$_{\rm eff}\simeq$1400--700 K;][]{golimowski04a,vrba04}.
The full catalogue of L and T dwarfs is available on the DwarfArchives.org
webpage\footnote{http://spider.ipac.caltech.edu/staff/davy/ARCHIVE/, 
a webpage dedicated to L and T dwarfs maintained by C.\ Gelino, 
D.\ Kirkpatrick, and A.\ Burgasser.}.
There are currently 46 T5 or later dwarfs and 17 T7--T8 dwarfs known
(two of them are marked as peculiar), at the
time of writing. The spectral classification of T dwarfs follows the 
unified scheme by \citet{burgasser06a} and is based on the strength 
of methane and water absorption bands present in the near-infrared. 
This sample of ultracool dwarfs is now
large enough to characterise the binary properties of brown
dwarfs \citep{close02b,burgasser03a,bouy03,burgasser06d,liu06,burgasser07d} 
and investigate the influence of gravity and metallicity on their 
spectral energy distributions \citep{kirkpatrick05}.

The UKIRT Infrared Deep Sky Survey \cite[UKIDSS;][]{lawrence06} is
a new infrared survey conducted with the UKIRT Wide Field Camera
(WFCAM). The survey is now well underway with (at the time of
writing) three ESO-wide releases: 
the Early Data Release (EDR) in February 2006 \citep{dye06}, the Data 
Release 1 (DR1) in July 2006 \citep{warren07a} and the Data
Release 2 in March 2007 \citep{warren07b}. The Large Area Survey 
(LAS) will cover 4000 deg$^2$ in $YJHK$ down to a 5$\sigma$ 
sensitivity limit of $J \simeq$ 19.5 mag in each of two epochs
(J$\simeq$20 mag when the epochs are combined) with a typical baseline of 
$\sim$2yr.

The two major science drivers of the LAS are the discovery of extremely 
cool brown dwarfs (with spectral types even later than the coolest 
objects discovered in the Two Micron All Sky Survey (2MASS) and 
the Sloan Digital Sky Survey (SDSS)), and  the discovery of
high-redshift quasars ($z \geq$ 6). New record-breaking 
low-$T_{\rm eff}$ dwarfs could require a new spectroscopic class 
beyond T \citep[e.g.][]{burrows03} which has been pre-emptively 
called Y \citep[following][]{kirkpatrick99}. 
So far UKIDSS has succeeded on both fronts; \citet{venemans07} 
report the discovery of a $z = 5.9$ quasar, and the discovery 
of two T dwarfs in the EDR \citep{kendall07}
has been followed by the identification of 
the latest T-type brown dwarf yet found \citep{warren07c}.

In this paper we report the discovery of eight T4.5--T7.5 dwarfs 
extracted from 190 deg$^2$ released in the LAS DR1.
We also derive kinematic properties for SDSS J020742.91$+$000056.2 
(hereafter SDSS0207), a T4.5 dwarf discovered by \citet{geballe02} and 
recovered in our search. In Section \ref{dT:selection} we describe 
the photometric selection of late-T dwarfs from two-colour diagrams.
In Section \ref{dT:spectro} we present the spectroscopic follow-up 
observations conducted with the Gemini and UKIRT telescopes, and classify 
each object based on the  \citet{burgasser06a} scheme.
In Section 4 we discuss the number of T dwarfs that our study suggests 
will be discovered in the complete LAS, and compare it with the 
predictions of \citet*{deacon06}. Finally, we summarise our results 
and give our conclusions in Section \ref{dT:conclusions}.

%
%%%%%%%%%%%%%%%%%%%%%%%%%%%%%%%%%%%%%%%%%%
%%%%%%%%%% Candidate selection %%%%%%%%%
%%%%%%%%%%%%%%%%%%%%%%%%%%%%%%%%%%%%%%%%%%
%
\section{Selection of T dwarf candidates}
\label{dT:selection}

\begin{figure}
  \includegraphics[width=\linewidth]{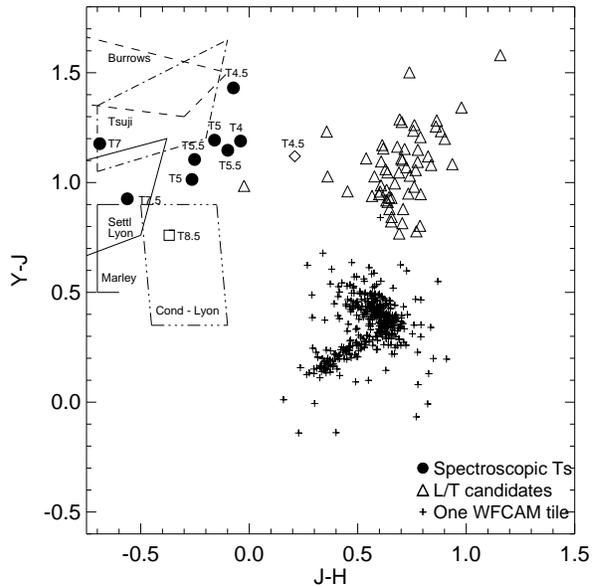}
   \caption{($J-H$, $Y-J$) two-colour diagram for point sources 
(small crosses) in one WFCAM tile centered on 
(RA,dec)=(02$^{\rm h}$,00$^{\circ}$).
Overplotted are L and T dwarfs candidates extracted from the
LAS DR1 whose spectroscopy is currently underway (open triangles).
Spectroscopically-confirmed T4.0--T7.5 dwarfs from DR1 presented 
in this paper are displayed as filled circles, and the T4.5 dwarf
from the EDR reported by \citet{kendall07} is shown as a diamond.
Note that two T5.5 dwarfs have the same colours and thus the circles
overlap. The T8.5 dwarf marked as an open square was selected from
DR1 and is presented in \citet{warren07c}.
Errors in the colours are typically better than 0.15 mag
for the faintest objects. Also shown are the  model
predictions for the colours of cool brown dwarfs from
the Cond \citep[dot-dot-dot-dashed lines;][]{allard01,baraffe03},
and Settl models (solid lines; Allard et al.\ 2007), and from models 
by Marley and collaborators \citep[solid lines;][]{marley02}, 
\citet[][dashed lines]{burrows03}, and 
\citet[][dot-dashed lines]{tsuji04}.
}
   \label{fig_dT:2col}
\end{figure}

This section describes the photometric search carried out
for late-T and possible Y dwarf candidates in the LAS, which is based on 
our current knowledge of the photometric properties of T dwarfs
from 2MASS and SDSS, as well as on  theoretical models.
%
%\subsection{Expected T and Y colours}
%\subsection{Expected brown dwarf colours}
%
\label{dT:context}
In addition to the traditional $JHK$ filters, a new $Y$ filter centered 
at 1.03 (0.98--1.08) microns was specifically designed and installed 
in WFCAM to ease the selection and separation of high-redshift 
quasars and cool brown dwarfs in the LAS colour-colour
diagrams \citep*{warren02}. Figure \ref{fig_dT:2col} shows ($Y-J$,$J-H$) 
for candidate and confirmed L and T dwarfs found in DR1, as well as 
model-predicted typical colours, and typical colours of point sources 
in a WFCAM tile.

Known T dwarfs found by 2MASS exhibit neutral to blue near-infrared
colours with decreasing effective temperature \citep{burgasser02} 
and a rather constant $Y-J\sim1$ \citep{hewett06}. SDSS 
discoveries \citep[e.g.][]{geballe02} show that T dwarfs are
red in the optical ($i-z\geq2.0$) and optical-to-infrared 
($z-J \geq 2.5$) colours.
the latter also being a good indicator of $T_{\rm eff}$ \citep{knapp04}. 
Hence the combination 
of the SDSS and the UKIDSS LAS should produce T and possibly Y dwarfs 
that are several magnitudes fainter than the 2MASS completeness limit. 

Current atmosphere models predict blue near-infrared colours 
($J-H < 0.0$ and $J-K < 0.0$) for dwarfs cooler than the known T dwarfs, 
but differ on the $Y-J$ optical-to-infrared colour: 
the cloud-free Cond models \citep{allard01,baraffe03},
%from the Lyon group 
the more recent Settl models (Allard et al.\ 2007, in prep.)
and the models of Marley et al.\ (2002)
all imply $Y-J$ colours that are bluer than those of the 
known late-T dwarfs, for $T_{\rm eff}\sim$ 700--400 K. 
However, \citet{burrows03} and \citet{tsuji04} predict
redder $Y-J$ colours for these temperatures \citep{leggett05,hewett06}.
Our initial search focussed on the redder $Y - J$ sources,
and a subsequent search for bluer objects yielded only one 
very late T dwarf, presented in a separate paper \citep{warren07c}.

%
%%%%%%%%%%%%%%%%%%%%%%%%%%%%%%%%%%%%%%%%%%%
%%%%%% Figure: 2-colour diagram %%%%%%%%
%%%%%%%%%%%%%%%%%%%%%%%%%%%%%%%%%%%%%%%%%%%
%
% Plot made with the IDL program ukidss_LAS_DR1_jhyj.pro
%
%

%
%%%%%%%%%%%%%%%%%%%%%%%%%%%%%%%%%%%%%%%%%%
%%%%%%%%%%%%% TABLE OBJECT %%%%%%%%%%%%%
%%%%%%%%%%%%%%%%%%%%%%%%%%%%%%%%%%%%%%%%%%
%
% Main data for our target
%
\begin{table*}
 \centering
 \caption[]{List of the coordinates (J2000), infrared magnitudes
            ($YJHK$) on the WFCAM/MKO-system and their associated 
            uncertainties, for ten T dwarfs
            identified in the UKIDSS LAS DR1 (ordered by right ascension),
            including SDSS J020742.83$+$000056.2 \citep{geballe02}
            and ULAS J003402.77$-$005206.7 \citep{warren07c}. 
            The 5$\sigma$ limits are
            computed using sky noise following the prescription of
            \citet{dye06}. Two sets of $JHK$ observations are available for
            ULAS J223955.76$+$003252.6: the first one originates
            from the UKIDSS LAS DR1; the second one comes from
            additional UKIRT/UFTI observations taken 03 September 2006
            (Section \ref{dT:new_phot}).
            $z$-band photometry was also obtained for this source with EMMI at the
            NTT and we measured $z_{N}(AB)$ = 22.46$\pm$0.09 mag.
            The quoted errors do not include calibration uncertainties; 
            these may be as high as 0.1 mag at $Y$, and 0.02 mag
            at $JHK$ \citep{warren07a}.}
 \begin{tabular}{c c c c c c c}
 \hline
Name  &  RA   & dec   &   $Y$  &  $J$  &  $H$  & $K$  \cr
\hline
ULAS J002422.94$+$002247.9  & 00 24 22.94 & $+$00 22 47.9 & 19.59$\pm$0.15 & 18.16$\pm$0.07 & 18.24$\pm$0.16 & $>$18.05  \cr
ULAS J003402.77$-$005206.7  & 00 34 02.77 & $-$00 52 06.7 & 18.90$\pm$0.10 & 18.14$\pm$0.08 & 18.50$\pm$0.22 & $>$17.94  \cr
ULAS J020336.94$-$010231.1  & 02 03 36.94 & $-$01 02 31.1 & 19.06$\pm$0.10 & 18.04$\pm$0.05 & 18.31$\pm$0.12 & 18.13$\pm$0.17  \cr
SDSS J020742.91$+$000056.2  & 02 07 42.91 & $+$00 00 56.2 & 17.94$\pm$0.03 & 16.75$\pm$0.01 & 16.79$\pm$0.04 & 16.71$\pm$0.05  \cr
ULAS J082707.67$-$020408.2  & 08 27 07.67 & $-$02 04 08.2 & 18.29$\pm$0.05 & 17.19$\pm$0.02 & 17.44$\pm$0.05 & 17.52$\pm$0.11  \cr
ULAS J090116.23$-$030635.0  & 09 01 16.23 & $-$03 06 35.0 & 18.82$\pm$0.05 & 17.90$\pm$0.04 & 18.46$\pm$0.13 & $>$18.21  \cr
ULAS J094806.06$+$064805.0  & 09 48 06.06 & $+$06 48 05.0 & 20.03$\pm$0.14 & 18.85$\pm$0.07 & 19.46$\pm$0.22 & $>$18.62  \cr
ULAS J100759.90$-$010031.1  & 10 07 59.90 & $-$01 00 31.1 & 19.82$\pm$0.12 & 18.67$\pm$0.07 & 18.77$\pm$0.18 & $>$18.17  \cr
ULAS J101821.78$+$072547.1  & 10 18 21.78 & $+$07 25 47.1 & 18.90$\pm$0.08 & 17.71$\pm$0.04 & 17.87$\pm$0.07 & 18.12$\pm$0.17  \cr
ULAS J223955.76$+$003252.6  & 22 39 55.76 & $+$00 32 52.6 & 19.94$\pm$0.17 & 18.86$\pm$0.09 &  $>$18.87        &  $>$18.18         \cr
                            &             &               &          & 18.85$\pm$0.05$^a$ & 19.10$\pm$0.10$^a$ & 18.88$\pm$0.06$^a$  \cr
\hline
\multicolumn{7}{|l|}{$^a$ UKIRT/UFTI Photometry}
 \label{tab_dT:objects}
 \end{tabular}
\end{table*}
%

%
%%%%%%%%%%%%%%%%%%%%%%%%%%%%%%%%%%%%%%%%%%
%%%%%% Figure: Finding charts %%%%%%
%%%%%%%%%%%%%%%%%%%%%%%%%%%%%%%%%%%%%%%%%%
%
% Crop of MultiGetImage results
%
\begin{figure*}
   \centering
  \includegraphics[width=0.49\linewidth, angle=0]{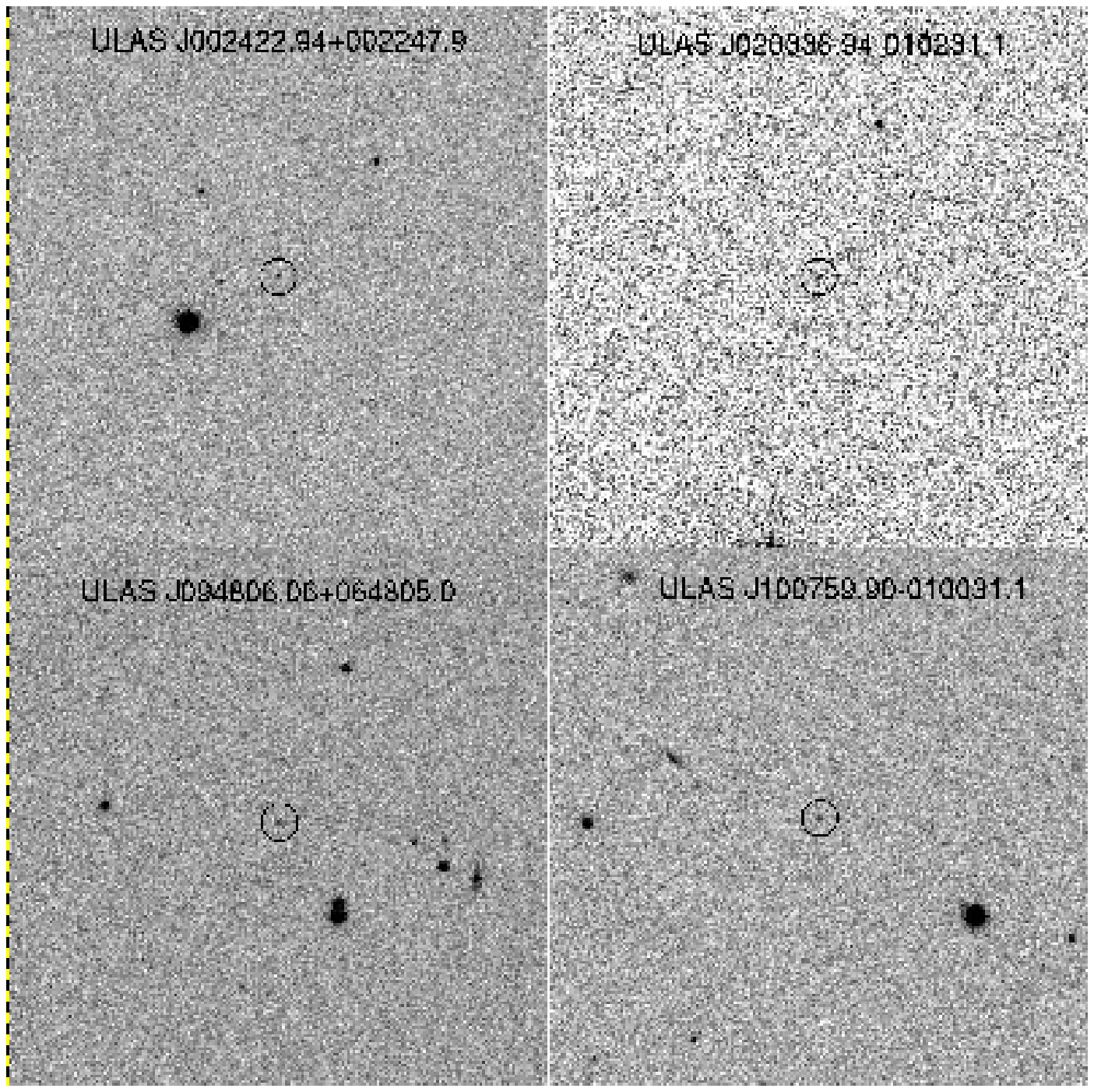}
  \includegraphics[width=0.49\linewidth, angle=0]{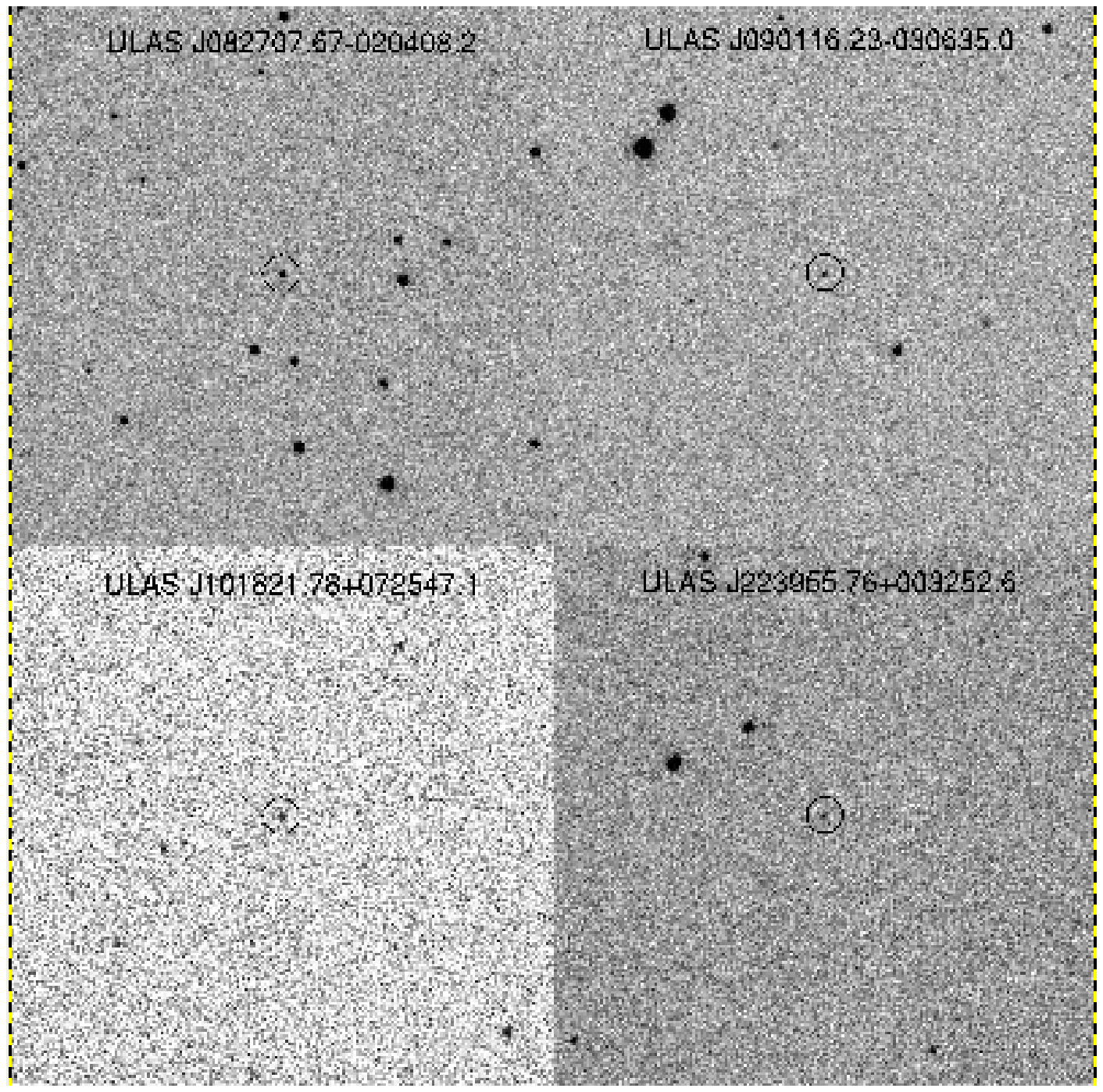}
   \caption{Finding charts for the eight new T4.5--T7.5 dwarfs 
extracted from the UKIDSS LAS DR1 and presented in this paper. 
Charts are $J$-band images of 2 arcmin on a side with North up and 
East left.
}
   \label{fig_dT:fcTs}
\end{figure*}
\subsection{Sample selection}
\label{dT:sample}

Our basic search methodology was to extract from the UKIDSS LAS DR1
a sample of point-sources with the $YJHK$ colours of late-T dwarfs, 
guided by the known and modelled colours, and then to cross-match with 
the SDSS (where possible) to obtain optical-to-infrared colours.

We have constructed a Structured Query Language (SQL) query to 
extract a list of reliable late-T dwarf candidates from the WFCAM 
Science Archive\footnote{located at http://surveys.roe.ac.uk/wsa/} 
(WSA; Hambly et al.\ 2007, in prep.). Further details regarding the
SQL statements are provided in Appendix \ref{dT:SQL_details}. 
We search for point
sources detected in at least $Y$ and $J$,
and use other post-processing flags to avoid galaxies and fast 
moving objects. In addition, we have limited our search to the
$J$ = 14--19 magnitude range, to avoid saturated sources and 
maximise the completeness of our
sample. Furthermore, we have imposed colour cuts of $Y-J \geq$ 0.5
and $J-H \leq$ 0.0 to focus on late-T dwarfs and potentially cooler
objects. Finally, images were checked to insure that the
candidates were real and unaffected by artefacts.
The location of the new T dwarfs in the ($J-H$,$Y-J$) two-colour
diagram is displayed in Fig.\ \ref{fig_dT:2col}.

For the 150 deg$^2$ of the UKIDSS LAS DR1 with SDSS data,
only those LAS-selected candidates which had $z - J \geq 2.5$, 
or were undetected by SDSS, were retained.  
Candidates in the second category were also subject to the 
further requirement that they were not within a few arcsec of 
a brighter point-source -- whilst such pairs are almost always 
separated in UKIDSS, the fainter source often remains unregistered 
in SDSS, leading to appreciable numbers of false $z$-band drop-outs.
Having applied all these automated filters, we were left with a total 
of 19 objects, 10 of which were quickly revealed to be spurious 
by visual inspection leaving a final list of just 9 viable 
T dwarf candidates. Of these, one (ULAS J022200.43$-$002410.5) 
is awaiting further investigation,
one is a known T4.5 dwarf (SDSS J020742.91$+$000056.2, 
Geballe et al.\ 2002), and one is a new UKIDSS T8.5 dwarf 
(ULAS J003402.77$-$005206.7, \citet{warren07c}).
The remaining six have been confirmed as new T dwarfs.
Finding charts are shown in Fig.\ \ref{fig_dT:fcTs},
their basic observational properties are summarised in 
Table \ref{tab_dT:objects}, and a brief statistical analysis 
of the sample is given in Section \ref{dT:density}.

For the 40 deg$^2$ of the UKIDSS LAS DR1 outside the SDSS DR5 footprint,
we rejected sources with optical counterparts
in the United States Naval Observatory (USNO) catalogues.
However the magnitude limit of these catalogues is so bright that
several hundred candidates remained, requiring 
cross-correlation with other optical catalogues or visual inspection.
Not all of these objects have been followed up,
but spectra were obtained of two of the most promising,
ULAS J082707.67$-$020408.2 and ULAS J090116.23$-$030635.0,
both of which were confirmed as T dwarfs.
Their basic parameters are listed in Table\ \ref{tab_dT:objects} 
and finding charts are given in Fig.\ \ref{fig_dT:fcTs}.
%but they are not included in the subsequent statistical 
%analysis (Section \ref{dT:density}) due to the slightly
%ad hoc selection process.

%
%%%%%%%% Additional photometry %%%%%%%%
%
\subsection{Additional photometry}
\label{dT:new_phot}

Additional optical and near-infrared photometry was obtained with
UKIRT and the New Technology Telescope (NTT) for one T dwarf, 
ULAS J223955.76$+$003252.6 (hereafter ULAS2239).
Details of these observations are given below.

ULAS2239 was observed at UKIRT
on UT 2006 September 03, using the UKIRT Fast-Track Imager
\citep[UFTI;][]{roche03}. The Mauna Kea Observatories
$J$, $H$ and $K$ filters were used \citep*[MKO;][]{tokunaga02},
and the data calibrated using UKIRT Faint Standards; the photometric
system should be identical to the WFCAM system \citep{leggett06}.
The night was photometric with 0.6 arcsec seeing. Individual
exposure times on target were one minute, and the total exposure time
was nine minutes at $J$, 18 minutes at $H$ and 27 minutes at $K$, 
dithering the dwarf on a 3$\times$3 grid with 10 arcsec offsets.
Photometry was measured using a two arcsec diameter aperture and
is presented in Table \ref{tab_dT:objects}.

In addition, we obtained $z$-band photometry (filter $\sharp611$)
for ULAS2239 with the ESO Multi-Mode Instrument (EMMI), at the NTT, 
on the night beginning 2006 August 18. 
The total exposure time was 15 minutes. The filter/detector 
combination is referred to here as
$z_N$ and is similar to the SDSS $z$. Using the measured CCD
sensitivity curve, and the filter transmission curve, we followed the
procedure used by \citet{hewett06} to establish the colour relation
$z_N(AB) = z(AB)-0.05(i(AB) - z(AB))$, for O to M dwarf stars. The offset
to the Vega system is $z_N(AB) =z_N + 0.54$. Photometry in $i$ and $z$
of SDSS DR5 sources in the field was converted to $z_N$, and the
brightness of the source was measured by relative photometry using a
fixed aperture, with the result $z_N(AB) = 22.46\pm0.09$. For a T6
dwarf $z(AB) \sim z_N(AB) + 0.2$.

%
%%%%%%%%%%%%%%%%%%%%%%%%%%%%%%%%%%%%%%%%%%
%%%%%% Figure: T5-T7 dwarf %%%%%%
%%%%%%%%%%%%%%%%%%%%%%%%%%%%%%%%%%%%%%%%%%
%
% Plot made with the program ukidss_DR1_GNIRSspec.pro
%
\begin{figure}
  \includegraphics[width=\linewidth]{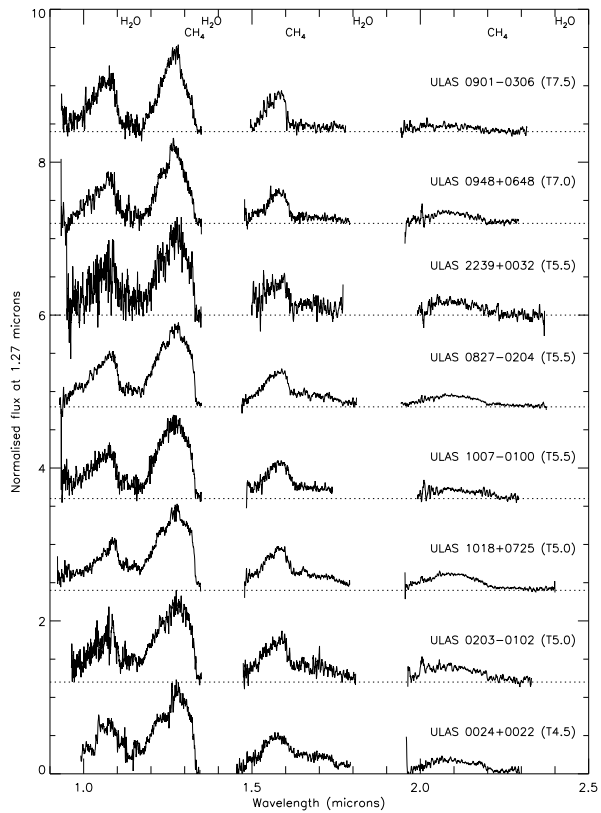}
   \caption{Gemini/GNIRS near-infrared (0.9--2.5 microns) spectra
of T4.5-T7.5 dwarfs confirmed spectroscopically.
From bottom to top ordered by increasing spectral type are:
ULAS J002422.94$+$002247.9 (T4.5), ULAS J020336.94$-$010231.1 (T5),
ULAS J101821.78$+$072547.1 (T5), ULAS J100759.18$-$010031.1 (T5.5),
ULAS J082707.67$-$020408.2 (T5.5), ULAS J223955.76$+$003252.6 (T5.5),
ULAS J094806.06$+$064805.0 (T7), and ULAS J090116.23$-$030635.0 (T7.5).
Spectra are smoothed by a boxcar of 10 pixels and shifted up
by increments of 1.2 for clarity (zero flux levels are indicated by the horizontal dotted lines).
Regions badly affected by telluric features (1.350--1.445 
and 1.811--1.942 microns) have been omitted.
}
   \label{fig_dT:T5_T7}
\end{figure}
%

%
%%%%%%%%%%%%%%%%%%%%%%%%%%%%%%%%%%%%%%%%%%
%%%%%% Table: Spectral types %%%%%%
%%%%%%%%%%%%%%%%%%%%%%%%%%%%%%%%%%%%%%%%%%
%
% Additional photometry from UKIRT
%
\begin{table*}
 \centering
 \caption[]{List of spectral types derived from indices 
            and from the comparison with T dwarf templates
            (c.f. \citet{burgasser06a}).
            The CH$_{4}$-$K$ index derived for the latest T dwarf is
            very uncertain due to low signal-to-noise in the $K$-band.}
 \begin{tabular}{c c c c c c c c}
 \hline
Name                       & H$_{2}$O-$J$ & CH$_{4}$-$J$ & H$_{2}$O-$H$ & CH$_{4}$-$H$ & CH$_{4}$-$K$  & Template & Adopted      \cr
\hline
ULAS J002422.94$+$002247.9 & 0.291 (T4.5) & 0.382 (T5.5) & 0.423 (T3.5) & 0.547 (T4)   &  0.195 (T5.5) & T4.5     & T4.5$\pm$0.5 \cr
ULAS J003402.77$-$005206.7 & 0.012 (T8.5) & 0.144 (T8.5) & 0.133 (T8.5) & 0.096 (T8)   &  0.091 ($\geq$T7) & T8--T8.5 & T8.5$\pm$0.5 \cr
ULAS J020336.94$-$010231.1 & 0.288 (T4.5) & 0.429 (T4.5) & 0.328 (T5.5) & 0.410 (T5)   &  0.245 (T4.5) & T4.5--T5 & T5.0$\pm$0.5 \cr
SDSS J020742.91$+$000056.2 &  ---  (---)  &  ---  (---)  & 0.391 (T4)   & 0.565 (T4)   &  0.282 (T4)   & T4.0     & T4.0$\pm$0.5 \cr
ULAS J082707.67$-$020408.2 & 0.213 (T5.5) & 0.384 (T5.5) & 0.300 (T6)   & 0.384 (T5.5) &  0.238 (T4.5) & T5--T5.5 & T5.5$\pm$0.5 \cr
ULAS J090116.23$-$030635.0 & 0.075 (T7.5) & 0.226 (T7)   & 0.118 (T8)   & 0.186 (T7)   &  0.198 (T5)   & T7--T7.5 & T7.5$\pm$0.5 \cr
ULAS J094806.06$+$064805.0 & 0.087 (T7)   & 0.215 (T7.5) & 0.254 (T6.5) & 0.222 (T7)   &  0.115 (T6.5) & T7.0     & T7.0$\pm$0.5 \cr
ULAS J100759.90$-$010031.1 & 0.177 (T5.5) & 0.321 (T6)   & 0.303 (T6)   & 0.345 (T5.5) &  0.191 (T5.5) & T5.5     & T5.5$\pm$0.5 \cr
ULAS J101821.78$+$072547.1 & 0.304 (T4.5) & 0.437 (T4.5) & 0.376 (T4.5) & 0.451 (T4.5) &  0.151 (T6)   & T5.0     & T5.0$\pm$0.5 \cr
ULAS J223955.76$+$003252.6 & 0.195 (T5.5) & 0.324 (T6)   & 0.258 (T6.5) & 0.356 (T5.5) &  0.187 (T5.5) & T5.5--T6 & T5.5$\pm$0.5 \cr
\hline
 \label{tab_dT:SpT}
 \end{tabular}
\end{table*}
%

%
%%%%%%%%%%%%%%%%%%%%%%%%%%%%%%%%%%%%%%%%%%
%%%%%%%% Spectroscopic follow-up %%%%%%%
%%%%%%%%%%%%%%%%%%%%%%%%%%%%%%%%%%%%%%%%%%
%
%
\section{Spectroscopic follow-up}
\label{dT:spectro}
\subsection{Gemini/GNIRS spectroscopy}

The Gemini Near-Infrared Spectrograph \citep[GNIRS;][]{elias06a}
on Gemini South was used to make quick response observations, through
programme GS-2006B-Q-36\@. All photometric candidates were confirmed
as T dwarfs with spectral types later than T4\@.
GNIRS was used in cross-dispersed mode with the
32l/mm grism, the 1.0 arcsec slit (position angle of 0$^\circ$) and
the short camera, to obtain 0.9--2.5$\mu$m R$\sim$500 (per resolution
element) spectra. Triggered observations were made on the nights of 
2006 August 26, October 12, 20, 27, and 2007 January 04, 20
with total integration time for each observation of 16 minutes. 
In each case the target was nodded three arcsec along the
slit in an ``ABBA'' pattern using individual exposure times of 240\,s.
Calibrations were achieved using lamps in the on-telescope calibration
unit. A0 and early F stars were observed as spectroscopic standards, 
either directly before or after the target observations, at an airmass 
that closely matched the mid-point airmass of the target in order to 
remove the effects of telluric absorption. The observing
conditions included some patchy cloud (i.e.\ cloud cover$<$70 per 
cent-ile), seeing from 0.5--1.4 arcsec, and humidity ranging from 
10--50\%.

Data reduction was initially done using tasks in the Gemini GNIRS IRAF
Package. Files were prepared and corrected for offset bias using NSPREPARE
and NVNOISE, and order separation achieved with NSCUT. Each order was then
median stacked at the A and B positions, and a difference image obtained
using GEMARITH. Flat-field correction was not necessary since variations
across the six arcsec slit are less than 0.1\%. S-distortion correction 
and wavelength calibration were performed interactively using the telluric
star spectra and Argon arc lamp spectra, with NSAPPWAVE, NSSDIST and
NSWAVE. Further reduction was carried out using custom IDL
routines. Apertures ($\sim$1.5 arcsec wide) were centred on the
spectra at the A and B positions, and the sky residuals were fit (and
subtracted) using a surface constructed via a series of least-squares
linear fits across the slit (excluding pixels within the apertures),
with one fit for each spatial pixel row. Spectra were then extracted
by summing within the A and B position apertures and combining. Noise
spectra were also determined using a combination of the sky noise
(between the A and B position spectra), and the photon noise associated
with the spectra themselves. The target spectra were flux-calibrated 
on a relative scale using the telluric
standard spectra (after appropriate interpolation across any hydrogen lines in the standards)
with an assumed black body function for $T_{\rm eff}$=10000\,K
and 7000\,K for A0 and early F tellurics respectively. The spectral orders
were then trimmed of their noisiest portions, and the spectra normalised
to unity at 1.27 microns (the average level from 1.265--1.275 microns).
Although the shape of each spectrum will be correct, the absolute flux 
calibration is best determined by scaling the spectra to match the 
WFCAM photometry. We find that a single scaling factor reproduces the 
observed $JHK$ magnitudes within the 2--10\% WFCAM photometric 
uncertainties, except for the $K$ magnitudes of 
ULAS J020336.94$-$010231.1 and ULAS J101821.78$+$072547.1 which the 
spectra suggest are 1.5 and 2.5$\sigma$ brighter than the 
WFCAM measurements, respectively.
The final smoothed spectra of the eight new T dwarfs observed with GNIRS 
are displayed in Fig.\ \ref{fig_dT:T5_T7}.

\subsection{UKIRT/UIST spectroscopy}

\begin{figure}
  \includegraphics[width=\linewidth]{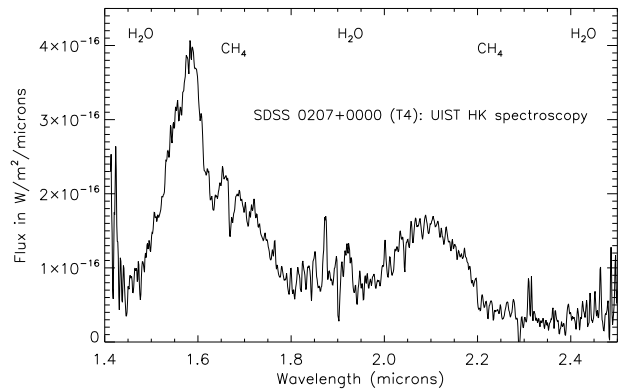}
   \caption{UKIRT/UIST near-infrared spectrum 
($HK$ grism; 1.4--2.5 microns) of a bright T4.0 dwarf (SDSS0207) 
selected photometrically from the UKIDSS LAS DR1 and confirmed
spectroscopically. The discovery 1.0--2.5 microns spectrum is 
presented in \citet{geballe02}.
}
   \label{fig_dT:T4dwarf}
\end{figure}

SDSS0207 was observed at UKIRT 
on UT 2006 September 04, using the UKIRT Imager Spectrometer 
\citep[UIST;][]{ramsay-howat04}. The HK grism was used with the four pixel 
slit, giving a resolution R=550\@. Individual exposure times on target 
were 240\,s, and the total exposure was 64 minutes, nodding the 
target along the slit by 12 arcsec. The instrument calibration lamps 
were used to provide accurate flat-fielding and wavelength calibration. 
The F5V star HD\,5892 was observed prior to the target to remove the 
effects of telluric absorption, and to provide an approximate flux 
calibration. More accurate absolute flux calibration was achieved by 
scaling the spectra to the $H$ and $K$ magnitudes measured in the LAS\@.
The final spectrum of SDSS0207 is shown in Fig.\ \ref{fig_dT:T4dwarf}.

%
%%%%%%%%%%%%%%%%%%%%%%%%%%%%%%%%%%%%%%%%%%
%%%%%% Figure: T4 dwarf %%%%%%
%%%%%%%%%%%%%%%%%%%%%%%%%%%%%%%%%%%%%%%%%%
%
% Plot made with the program ukidss_DR1_GNIRSspec.pro
%
\subsection{Spectral classification}
\label{dT:classification}

For each spectrum we calculated the five near-infrared spectral 
indices H$_2$O-$J$, CH$_4$-$J$, H$_2$O-$H$, CH$_4$-$H$ and 
CH$_4$-$K$ as defined in \citet{burgasser06a}. We also compared 
the data by eye to template spectra for T dwarfs defined by 
Burgasser et al. as standards for the T4, T5, T6, T7 and T8 types. 
The results for the new T dwarfs with GNIRS spectra are given in 
Table \ref{tab_dT:SpT}. For the recovered SDSS0207, we derive 
a type of T4.0$\pm$0.5, in agreement with \citet{geballe02} and 
\citet{burgasser06a} who derive T4.5$\pm$0.5.
Some scatter is present in the spectral indices given in
Table \ref{tab_dT:SpT}. The GNIRS exposures were designed to provide 
only a rough spectral type for potential late-T dwarfs, and so can
yield somewhat noisy spectra, especially in the $K$-band.
Consequentially, the direct comparison with templates has been given
more weight than the spectral indices in the assignment of spectral
types, and the adopted uncertainty in Table \ref{tab_dT:SpT} reflects 
the range in type implied by this comparison.

The comparison to the spectral templates, together with modelled 
trends with gravity and metallicity (as shown for example by \cite{liu07}, 
and Leggett et al. 2007, ApJ, submitted) allow us to identify dwarfs that 
may have a metallicity or a gravity  
different from the solar-neighbourhood sample that defines the templates.
% ADD REFS
% Liu, Michael C.; Leggett, S. K.; Chiu, Kuenley 2007 ApJ in press
% Leggett, S. K. et al., 2007, ApJ submitted
Leggett et al. (2007, ApJ, submitted; their Figure 3) shows that, 
for late-T dwarfs, increasing gravity 
suppresses the $K$-band flux due to increased pressure-induced H$_2$ 
opacity (and vice-versa), while decreasing metallicity also increases the 
H$_2$ opacity and suppresses $K$, but has the additional signature 
of broadening the 1 micron $Y$-band flux peak. Searching for these 
signatures, we find that most of our sample appear to be typical 
of the solar-neighbourhood, so that metallicity is likely to be solar, 
and gravity is likely to be given by $\log g$=5.0. The exceptions are 
(see Fig.\ \ref{fig_dT:T5_T7}): ULAS J020336.94$-$010231.1 which 
appears to be a low-gravity object based on the high $K$-band flux; 
ULAS J090116.23$-$030635.0 which appears to be a high-gravity object 
based on the low $K$-band flux; and ULAS J101821.78$+$072547.1 
which appears to be metal-rich based on the
high $K$-band flux and narrow $Y$-band flux peak.
ULAS J223955.76$+$003252.6 may have a somewhat low gravity, however
the spectrum is very noisy for this very faint dwarf. 
Fig. \ \ref{fig_dT:T5_T7} also suggests that the $K$-band flux of 
ULAS J100759.90$-$010031.1 is suppressed, but the spectrum is 
compromised by data spikes at 2.0 and 2.2 microns, and
the spectral energy distribution is in fact very similar to the 
templates if these regions are ignored.

%
%%%%%%%%%%%%%%%%%%%%%%%%%%%%%%%%%%%%%
%%%%%%%  Discussion %%%%%%%
%%%%%%%%%%%%%%%%%%%%%%%%%%%%%%%%%%%%%
%
\section{Discussion}
\label{dT:discussion}
\subsection{Properties of SDSS J020742.83$+$000056.2}
\label{dT:kinematics}

Our search criteria recovered one T dwarf found by the SDSS group
\citep[SDSS0207 --- T4.5;][]{geballe02}. \citet{geballe02} 
reported $JHK$ magnitudes for SDSS0207 of 16.63, 16.66 and 16.62, 
with uncertainties of 0.05 mag. Observations were obtained with 
UKIRT/UFTI and are described in \citet{leggett02}. 
Although the WFCAM magnitudes (Table \ref{tab_dT:objects}) are 
$\sim$0.1 magnitudes fainter, the two datasets agree to 2$\sigma$ 
and are thus reasonably consistent. The second epoch LAS could 
be used to investigate photometric variability in the T dwarfs.
The derived spectral types (T4.5 in the discovery paper and T4
here are in agreement within the classification uncertainties.

This object has known proper motion (0.156$\pm$0.011 arcsec/yr)
and parallax \citep[0.035$\pm$0.001 arcsec;][]{vrba04}. We are able
to estimate its proper motion from the $\sim$3 year baseline between
the SDSS (2002 September 04) and LAS (2005 November 26) observations. 
We have measured 0.142 and 0.085 arcsec/yr in right ascension
and declination, respectively, yielding a total proper motion
of 0.166$\pm$0.020 arcsec/yr consistent with the value given 
in \citet{vrba04}.
SDSS observations can thus provide a first epoch comparison
for UKIDSS, improving the prospects for identifying sources with 
significant proper motion in the second-epoch $J$-band LAS observations.

\subsection{Distances}
\label{dT:distances}
%
% Three estimates are available: from Vrba et al. (2004) using
% parallaxes, from Knapp et al. (2004) using a 5-order polynomial
% fit and from Liu et al. (2006) using the same type of fit
% but excluding known and possible binaries.
%
% ulas0024  T4.5  J=18.162  d=
% ulas0203  T5.0  J=18.040  d=47.86 (38.02-60.25)
% ulas1018  T5.0  J=17.710  d=
% ulas0827  T5.5  J=17.190  d=29.38 (23.33-36.98)
% ulas1007  T5.5  J=18.672  d=
% ulas2239  T5.5  J=18.850  d=63.10 (50.12-79.43)
% ulas0948  T7.0  J=18.853  d=
% ulas0901  T7.5  J=17.900  d=22.49 (17.86-28.31) less than 25 pc
%

To estimate the distances of the new T dwarfs we considered three
relationships between spectral type and absolute magnitude: those 
published by \citet{vrba04}, \citet{knapp04}, and \citet{liu06}. 
The first was derived using 19 T0--T8 dwarfs with measured parallaxes. 
\citet{knapp04} supplied a fifth-order polynomial fit to 42 
spectroscopically confirmed T dwarfs. \citet{liu06} 
employed the same procedure as \citet{knapp04} but excluded known 
and possible binaries from the fit to take into account the higher 
frequency of binary systems among objects with L/T transition
colours \citep[e.g.][]{burgasser06d,liu06}. The exclusion of binaries 
results in a relation with fainter absolute magnitudes and smaller 
inferred distances, for early-mid T dwarfs. The values derived using 
the three relationships are given in Table \ref{tab_dT:Dist}.

For our final distance estimates we chose those derived using 
the most recent spectral-type--M$_J$ relation \citep{liu06}.
The uncertainties in our distance estimates were computed by assuming 
a dispersion of $\pm$0.5 mag in the spectral type--absolute magnitude 
relation. The adopted mean distances and their associated intervals
are quoted in Table \ref{tab_dT:Dist}. These distances assume single
objects; if any of these sources are in fact multiple systems, 
they would be more distant by 40\%, assuming an unresolved 
binary with components of comparable brightness.

These spectroscopic distances show that the LAS is detecting 
late-T dwarfs out to $\sim$80\,pc \citep[see also][the
discovery of ULAS J1452$+$0655, a T4.5 dwarf at around 80\,pc,
shown in our Figure 1 as a diamond]{kendall07}.
Also, note that the T7.5 dwarf ULAS J090116.23$-$030635.0, 
and the T8.5 dwarf ULAS J003402.77$-$005206.7 \citep{warren07c}, 
may lie within 
the 25\,pc limit of the catalogue of nearby stars \citep{gliese95}, 
and thus represent an important addition to the list of nearby 
cool brown dwarfs.

Assuming the same spectral type-absolute magnitude relationship, 
we derive a spectroscopic distance of 28\,pc (23--36\,pc) 
for SDSS0207, in good agreement with the parallax distance of 
28$\pm$0.8\,pc \citep{vrba04}. This object is of interest 
as it populates the bump seen for the L-T sequence in the spectral 
type--M$_J$ diagram for late-L to mid-T dwarfs \citep{burgasser06d,liu06}. 
Although this bump is known to be partially 
caused by unresolved binarity, it also appears to be intrinsic
to the atmospheric physics of the L-T transition.

%
%%%%%%%%%%%%%%%%%%%%%%%%%%%%%%%%%%%%%%%%%%
%%%%%% Table: Distances %%%%%%
%%%%%%%%%%%%%%%%%%%%%%%%%%%%%%%%%%%%%%%%%%
%
% Distances for the 4 new late-T dwarfs
%
\begin{table}
 \centering
 \caption[]{Distances (pc) for the eight new late-T dwarfs presented 
            in this paper, based on relationships
            between absolute-magnitude and spectral type from
             \citet[][; V04]{vrba04}, \cite[][; K04]{knapp04}, 
            and \citet[][; L06]{liu06}. The adopted distances allow 
            for a 0.5 mag uncertainty on the fifth-order
            polynomial provided in \citet{liu06}. The distance for 
            the T8.5 ULAS J003402.77$-$005206.7 is taken from 
            \citet{warren07c} and is based on model spectral fits.}
 \begin{tabular}{@{\hspace{1mm}}c c c c c@{\hspace{1mm}}}
 \hline
Name  &   V04  &  K04  &  L06 &  Adopted \cr
\hline
ULAS J002422.94$+$002247.9 & 54.4 & 62.4 &  53.1 & 42--67  \cr
ULAS J003402.77$-$005206.7 &      &      &       & 14--21  \cr
ULAS J020336.94$-$010231.1 & 47.6 & 54.5 &  48.3 & 38--61  \cr
ULAS J082707.67$-$020408.2 & 29.2 & 33.1 &  30.8 & 24--39  \cr
ULAS J090116.23$-$030635.0 & 22.3 & 28.3 &  26.0 & 21--33  \cr
ULAS J094806.06$+$064805.0 & 41.4 & 41.4 &  47.7 & 38--60  \cr
ULAS J100759.90$-$010031.1 & 57.9 & 65.6 &  60.9 & 48--77  \cr
ULAS J101821.78$+$072547.1 & 40.9 & 46.6 &  41.5 & 33--52  \cr
ULAS J223955.76$+$003252.6 & 62.8 & 71.5 &  66.1 & 52--83  \cr
\hline
 \label{tab_dT:Dist}
 \end{tabular}
\end{table}
\subsection{Expected numbers of late-T dwarfs in the LAS}
\label{dT:density}

In this section we try to estimate the observed number density of
late-T ($\geq$T4; T$_{\rm eff} \leq$ 1300 K) dwarfs from our spectroscopic 
follow-up and compare it with predictions by \citet{deacon06}.
Quantification of completeness is difficult at this early stage
at the faint end of the survey. However, we expect our sample
to be complete down to $J$ = 18.5 mag and possibly $J$ = 19 mag
because all late-T dwarfs will be detected in the $Y$ passband.

This paper reports ten T dwarfs, with spectral type T4 and later, 
extracted from a total area of 190 deg$^2$ (DR1). Of these, 
seven are brighter than $J$ = 18.5 mag.  To compare directly 
with the numbers quoted in Table 1 of \citet{deacon06}, 
we need to scale our numbers to 4000 deg$^2$ (the full LAS coverage 
after seven years of operation) and take into account the final depth 
of the survey ($J \sim$ 20 mag) after two epochs. We assume 
here a uniform and constant distribution of ultracool dwarfs which 
is reasonable, because an extra magnitude of depth for  late-T dwarfs 
will not probe the scale height of the disk.

Consequently, for the bright sample with $J \leq$ 18.5 mag,
we should find 7$\times$4000/190
multiplied by a factor of 8 to account for the volume difference,
yielding a total of 1179$\pm$34 late-T dwarfs. If we take into account
all T dwarfs reported here, we estimate a total of 
(10$\times$4000/190)$\times$4 = 842$\pm$29 T dwarfs in the full LAS\@.
These values match the predicted numbers from the simulations of
\citet{deacon06} for a mass function with a slope between 
$\alpha$ = $-$1.0 and $\alpha$ = $-$0.5 (in Salpeter units or 
$\xi(\log m)$ = d$n$/d$\log m$ $\propto$ $m^{-\alpha}$; 
676--1060 late-T dwarfs, defined in \citet{deacon06} as
dwarfs with 1300 $\geq T_{\rm eff}$ (K) $\geq 700$). 
This result is also in agreement with the 
derivations of $\alpha$ = $-$0.7$\pm$0.7 for M $\leq$ 0.08 M$_{\odot}$ 
by \citet{kroupa01b}
and conclusions drawn by \citet{allen05} from their Bayesian
approach ($\alpha$ = $-$0.6$\pm$0.7 for M $\leq$ 0.04--0.1 M$_{\odot}$).
Similar results are found for the low-mass end of the IMF in young
open clusters \citep[e.g.\ $\sigma$ Orionis;][]{gonzales_garcia06}.

However, uncertainties in the expected number of T dwarfs and 
the slope of the mass function remain large on both the observational 
and theoretical sides. Malmquist related biases have not been taken 
into consideration for example, but contribute significant uncertainty. 
For instance, earlier T dwarfs could be detected out to greater 
distance than some mid-late Ts, and our colour selections are 
poorly suited for identifying these. Furthermore, there will be 
a bias favouring unresolved binary systems, as they are brighter 
than single T dwarfs and detectable at greater distance. 
The Deacon \& Hambly numbers do not account for this bias, but 
it will be inherent in our sample. A more extensive spectroscopic 
follow-up of L and T dwarf candidates from the LAS is necessary 
to disentangle the various possible shapes of the IMF\@.

%
%%%%%%%%%%%%%%%%%%%%%%%%%%%%%%%%%%%%%
%%%%%%%  CONCLUSIONS %%%%%%%
%%%%%%%%%%%%%%%%%%%%%%%%%%%%%%%%%%%%%
%
\section{Conclusions}
\label{dT:conclusions}

We have presented the spectroscopic confirmation of eight new T4.5--T7.5 
dwarfs identified over the entire LAS area released in the first 
UKIDSS data release, together with the recovery of a T4.5 dwarf 
already discovered in the SDSS, and a T8.5 reported by \citep{warren07c}.
The spectral classification is based on the unified T-dwarf classification 
scheme of \citet{burgasser06a}. In addition, we have estimated 
distances for all of our sources using recent spectral 
type--absolute magnitude relationships. The two latest and 
coolest dwarfs among the new T dwarfs 
identified in the LAS DR1, the T7.5 and the T8.5, might be within 
the 25 pc limit of nearby stars. Although our spectroscopic follow-up 
is limited, our estimates of the number density of T dwarfs agree with 
other studies.

We have demonstrated the capabilities of the UKIDSS LAS
and expect to achieve the main scientific drivers set for the survey.
The LAS will be able to detect T dwarfs out to $\sim$80 pc
and L dwarfs much further, enabling the study of the scale height
of field brown dwarfs.
Also, the second-epoch coverage in $J$ will provide 
proper motions for a large number of candidate L and T dwarfs identified 
in the LAS, and allow us to identify fainter and more distant objects.

%
%%%%%%%%%%%%%%%%%%%%%%%%%%%%%%%%%%%%%
%%%%%%%  ACKNOWLEDGEMENTS %%%%%%%
%%%%%%%%%%%%%%%%%%%%%%%%%%%%%%%%%%%%%
%
\section*{Acknowledgments}

NL was a postdoctoral research associate funded by the UK PPARC
at the University of Leicester where part of this work was carried out.
SKL is supported by the Gemini Observatory, which is operated by the 
Association of Universities for Research in Astronomy, Inc., on behalf 
of the international Gemini partnership.
MCL acknowledges support for this work from NSF grants AST-0407441 
and AST-0507833 and an Alfred P. Sloan Research Fellowship.
This research has made use of the Simbad database of NASA's
Astrophysics Data System Bibliographic Services (ADS).
Research has benefitted from the M, L, and T dwarf compendium housed
at DwarfArchives.org and maintained by Chris Gelino, Davy Kirkpatrick,
and Adam Burgasser.
The United Kingdom Infrared Telescope is operated by the Joint 
Astronomy Centre on behalf of the U.K. Particle Physics and Astronomy 
Research Council. Based on observations obtained at the Gemini 
Observatory (program GS-2006B-Q-36), which is operated by the 
Association of Universities for Research in Astronomy, 
Inc., under a cooperative agreement with the NSF on behalf of the 
Gemini partnership: the National Science Foundation (United States), 
the Particle Physics and Astronomy Research Council (United Kingdom), 
the National Research Council (Canada), CONICYT (Chile), the Australian 
Research Council (Australia), CNPq (Brazil) and CONICET (Argentina).
The SDSS is managed by the Astrophysical Research Consortium 
for the Participating Institutions. The Participating Institutions 
are the American Museum of Natural History, Astrophysical Institute 
Potsdam, University of Basel, University of Cambridge, Case Western 
Reserve University, University of Chicago, Drexel University, Fermilab, 
the Institute for Advanced Study, the Japan Participation Group, Johns 
Hopkins University, the Joint Institute for Nuclear Astrophysics, the 
Kavli Institute for Particle Astrophysics and Cosmology, the Korean 
Scientist Group, the Chinese Academy of Sciences (LAMOST), Los Alamos 
National Laboratory, the Max-Planck-Institute for Astronomy (MPIA), 
the Max-Planck-Institute for Astrophysics (MPA), New Mexico State 
University, Ohio State University, University of Pittsburgh, University 
of Portsmouth, Princeton University, the United States Naval Observatory, 
and the University of Washington.

%
%%%%%%%%%%%%%%%%%%%%%%%%%%%%%%%%%%%%%%%%%%
%%%%%%%%  Bibliography  %%%%%%%%
%%%%%%%%%%%%%%%%%%%%%%%%%%%%%%%%%%%%%%%%%%
%
\bibliographystyle{mn2e}
\bibliography{../../AA/mnemonic,../../AA/biblio_old}
%

%
%%%%%%%%%%%%%%%%%%%%%%%%%%%%%%%%%%%%%%%%%%
%%%%%%%%  Appendix  %%%%%%%%
%%%%%%%%%%%%%%%%%%%%%%%%%%%%%%%%%%%%%%%%%%
%
\appendix

%
%%%%%%%%%%%%%%%%%%%%%%%%%%%%%%%%%%%%%%%%%%
%%%%  Detailed selection procedure  %%%%
%%%%%%%%%%%%%%%%%%%%%%%%%%%%%%%%%%%%%%%%%%
%
\section{Selection procedure for T dwarf candidates in the UKIDSS
Large Area Survey}
\label{dT:SQL_details}

The set of Structured Query Language (SQL) statements employed to
extract lists of reliable point-sources with the colours of late-T 
dwarfs from the WFCAM Science Archive is detailed here.
Sufficient reliability was obtained by demanding detection in at
least the $Y$- and $J$-bands
(\{{\tt{Y}},{\tt{J\_1}}\}{\tt{Class > -10}}),
rejecting sources which were classified as noise in any band
(\{{\tt{Y}},{\tt{J\_1}},{\tt{H}},{\tt{K}}\}{\tt{Class != 0}}, 
note that this is different from a null detection which has 
{\tt{Class = $-$9999}}), and requiring that inter-band
positional offsets are no more than 0.7 arcsec to preclude 
asteroids which have moved appreciably between exposures.
Point-sources were selected by demanding {\tt{MergedClass = -1}},
and we rejected bright, potentially saturated sources
({\tt{J\_1AperMag3 >= 14.0}} mag), in addition to applying a
magnitude cut of $J \leq 19$ mag.
The principal T dwarf selection was simply a pair of colour cuts:
$Y-J \geq 0.5$ and $J-H \leq 0.0$ ({\tt{YAperMag3 - J\_1AperMag3 >= 0.5}}
and {\tt{J\_1AperMag3 - HAperMag3 <= 0.0}}),
where, in the case of non-detections, the $H$-band magnitude
was replaced with the $5\sigma$ $H$-band point-source limit
calculated using the recipe given by \citet{dye06}.
The candidate list was further trimmed by removing potential
cross-talk artefacts \citep{dye06} by
rejecting any object up to seven multiples of 51 arcsec away
in right ascension or declination from a 2MASS point-source 
with $J \leq 13.5$ mag.

\label{lastpage}

\end{document}